\documentclass[runningheads]{llncs}
\usepackage{graphicx}
\usepackage{listings}
\usepackage[colorlinks=false]{hyperref}
\usepackage[T1]{fontenc}
\begin{document}
\title{SRTK: A Toolkit for Semantic-relevant Subgraph Retrieval}
%
%
\author{Yuanchun Shen \orcidID{0009-0003-1513-5362}}
\authorrunning{Yuanchun Shen}
%
\institute{Technical University of Munich\\
\email{y.c.shen@tum.de}}
\maketitle              
\begin{abstract}
Information retrieval based knowledge base question answering (KBQA) first retrieves a subgraph to reduce search space, then reasons on the subgraph to select answer entities. Existing approaches have three issues that impede the retrieval of such subgraphs. Firstly, there is no off-the-shelf toolkit for semantic-relevant subgraph retrieval. Secondly, existing methods are knowledge-graph-dependent, resulting in outdated knowledge graphs used even in recent studies. Thirdly, previous solutions fail to incorporate the best available techniques for entity linking or path expansion. In this paper, we present SRTK, a user-friendly toolkit for semantic-relevant subgraph retrieval from large-scale knowledge graphs. SRTK is the first toolkit that streamlines the entire lifecycle of subgraph retrieval across multiple knowledge graphs. Additionally, it comes with state-of-the-art subgraph retrieval algorithms, guaranteeing an up-to-date solution set out of the box.

\textbf{Resource Type}: Software \\
\textbf{License}: MIT \\
\textbf{DOI}: \url{https://doi.org/10.5281/zenodo.7895612} \\
\textbf{Repository}: \url{https://github.com/happen2me/subgraph-retrieval-toolkit}
\end{abstract}
\section{Introduction}

Knowledge base question answering (KBQA) aims to answer natural language questions over large scale knowledge graphs such as Wikidata \cite{vrandevcic2014wikidata}, Freebase \cite{bollacker2008freebase}, and DBPedia \cite{auer2007dbpedia}.  One crucial step in KBQA is subgraph retrieval, which involves narrowing down the search space by retrieving a subset of entities and relations from the knowledge graph that are relevant to the question. By obtaining a smaller subgraph with more pertinent information, noise can be reduced, and the reasoning process can be facilitated.

Subgraph retrieval can be divided into two main steps: entity linking and path expansion. Entity linking identifies named entities in questions and anchors them to corresponding entities in knowledge graphs, while path expansion selectively includes neighbor entities and relations to form the subgraph. The scale and complexity of knowledge graphs pose challenges for both steps. Firstly, accurate entity linking requires detecting potential entity mentions and disambiguating them to the correct entities in the knowledge graph. This task becomes challenging due to the large number of entities within the knowledge graph, where multiple entities may have similar names to those mentioned in the question. Once the entities are identified, selecting the most relevant paths for subgraph expansion is also non-trivial. Although numerous paths can be expanded from an entity, only a small subset of them are truly relevant. For instance, as of the time of writing, Wikidata has 3,278 entities connected to the United States as the subject of certain relations, exemplifying the vast number of potential paths to consider.

Existing KBQA works that involve subgraph retrieval have certain limitations. For entity linking, some recent works still rely on pure pattern matching, resulting in a significant number of irrelevant linked entities \cite{yasunaga2022dragon,liu2021kgbart,das2022kgqacase}.  In terms of path expansion, many existing approaches either blindly expand entities within one or two hops \cite{chaudhuri2021grounding}, which quickly becomes infeasible for large knowledge graphs, or filter expansion paths using non-trainable embedding similarity \cite{yasunaga2021qagnn}, which may not capture the most relevant paths. Zhang et al. proposed a path expansion method called SR \cite{zhang2022ruckbqa}, which achieved state-of-the-art results on specific KBQA datasets, but it assumes pre-linked entities and only works with the outdated knowledge graph Freebase.

To address these gaps, we propose SRTK, a toolkit designed to simplify the retrieval of semantic-relevant subgraphs from large-scale knowledge graphs. SRTK integrates multiple off-the-shelf entity linking tools with unified interfaces and implements the SR path expansion algorithm \cite{zhang2022ruckbqa} for both Freebase and up-to-date knowledge graphs such as Wikidata and DBpedia. Furthermore, we extend the SR algorithm to support contrastive losses and different base models during training, as well as varying beam width and search depth during inference. To our knowledge, SRTK is the first readily available toolkit for subgraph retrieval across multiple knowledge graphs. Its main features include:

\begin{itemize}
    \item Out-of-the-box Functionality: SRTK provides command-line tool and Python library for subgraph retrieval. It comes with documentation and tutorials for a quick and effortless start.
    \item Full Lifecycle Support for Subgraph Retrieval: SRTK streamlines the complete lifecycle of subgraph retrieval.  This includes not only retrieval itself (entity linking, path expansion), but also retrieval model training (data processing, scorer training, and retrieval evaluation).
    \item Multi-Knowledge Graph Support: SRTK provides support for Freebase, Wikidata, and DBpedia, employing unified access patterns. Furthermore, the toolkit can be extended to accommodate other knowledge graphs that have a SPARQL interface, thereby increasing its versatility.
    \item User-friendly Design: SRTK offers a user-friendly interface that is both intuitive and extensible. For instance, each step of the retrieval process can be executed with a single command; various steps are seamlessly connected using standardized JSONL files, ensuring a smooth workflow; extension to new knowledge graphs and entity linking tools can be accomplished by implementing standardized protocols, etc;.
    \item Inclusion of SOTA Algorithms: REL and DBpedia are among the best available off-the-shelf entity linking tools; For path expansion, we employed the SR algorithm proposed by Zhang et al \cite{zhang2022ruckbqa}, which achieves the state-of-the-art results on two Freebase KBQA datasets, ensuring path expansion is on par with the state-of-the-art methods.
    \item Huggingface Model Compatibility: SRTK supports training or evaluation with any language encoding models available on the Huggingface model hub.
    \item Interactive Visualization: Retrieved subgraphs can be visualized as interactive web pages, allowing users to explore and analyze the retrieved information in a user-friendly and intuitive manner.
\end{itemize}

The rest of the paper is structured as follows. We begin with a review of related works in section 2. Then we introduce the main functions and implementations of SRTK in section 3, exemplifying the usage of SRTK.  Subsequently, in section 4, we discuss the impact, positioning to the state of the art, limitations, and future development plans. Finally, we conclude the paper in section 5.

\section{Related Works}

\textbf{KBQA} answers natural language questions with entities or relations from a knowledge graph \cite{lan2022kgqasurvey}. Knowledge graph is a structured representation of knowledge. It usually represents real-world facts as \textit{(subject, predicate, object)} triples. Researchers have approached the problem of KBQA in different ways, which can be broadly categorized into two types. The first type is based on semantic parsing, where a graph query is constructed by filling patterns or end-to-end constructions to directly retrieve the targets \cite{yih2015semanticparsing,das2022kgqacase}. However, this method has limitations due to the difficulty in precisely interpreting human intentions and the fact that it excludes potentially helpful reasoning paths and connected nodes. The second type is based on information retrieval. It regards selecting targets as either binary classification or ranking over candidate entities \cite{fu2020kbqasurvey}. They usually start with entity linking that matches mentions in question to entities in knowledge graphs, then expands the subgraphs by including neighbors. A reasoner is then applied to select correct answers from the retrieved subgraphs \cite{zhang2022ruckbqa,chaudhuri2021grounding,yasunaga2021qagnn}. Our work mainly applies to this branch of KBQA solutions.

\bigskip

\noindent \textbf{Semantic-relevant Subgraph Retrieval} reduces the search space and noisy information for downstream reasoners by retrieving a pertinent subgraph that is likely to contain the target entities or relations,  leveraging the semantic information conveyed in a natural language question. This retrieval process is typically divided into entity linking and path expansion. In entity linking, it first performs named entity recognition (NER) to detect entity mentions in the questions, then employs entity disambiguation to link entities to corresponding knowledge graphs. While in path expansion, neighbor entities and relations to existing entities are retrieved to form the subgraphs. The assumption behind this approach is that the target entities are within proximity of mentioned entities in the questions. There are also works that do not follow this approach. For example, CLOCQ \cite{christmann2022beyondned} use a combination of features, such as lexical matching, question-relevance, coherency, and connectivity, to directly retrieve top-$k$ relevant entities or relations as the reduced search space. 

Earlier methods to subgraph retrieval, which are still adopted by recent works \cite{yasunaga2021qagnn,liu2021kgbart,das2022kgqacase}, usually starts with keywords matching to link entities to knowledge graphs, then include all entities within a certain hop. Such methods are suitable for small-scale knowledge graphs, but for large knowledge graphs, lexically matching mentions may include many unrelated entities. Including all $k$-hop entities may further lead to a size explosion of the retrieved graph. Based on this approach, some optimizations were proposed to shrink the subgraph size. QA-GNN \cite{yasunaga2021qagnn} uses GLOVE embedding distance as similarity metrics to filter out less similar entities and relations, but it is not trainable, thus incapable for situations where the expansion path is semantically dissimilar to the question itself.

Recent advancements in subgraph retrieval have demonstrated improvements in both entity linking and path expansion. Several entity linking tools, such as WAT \cite{piccinno2014wat}, REL \cite{vanHulst2020REL}, DBpedia Spotlight \cite{mendes2011dbpediaspotlight}, have been developed for different knowledge graphs, incorporating lexical, contextual, and semantic information. In terms of path expansion, recent works take semantic information of the retrieval paths into consideration. PullNet\cite{sun2019pullnet} assumes that entities are already linked. It iteratively \textit{pulls} neighboring entities or entities from a related corpus based on their probabilities predicted by a neural network, forming the subgraph. However,  PullNet lacks the separation between path expansion and subgraph reasoning. Building upon this, Zhang et al. proposed SR \cite{zhang2022ruckbqa}, which decouples path expansion from reasoning to enable a plug-and-play framework that generalizes to different reasoners. SR performs iterative path expansion by comparing the trained semantic similarity of the next expansion path with the question concatenated with the previously expanded paths. By leveraging subgraphs retrieved through SR, Zhang et al. achieved state-of-the-art results on WebQSP \cite{yih2016webqsp} and CWQ \cite{talmor2018cwq} KBQA datasets. However, it is important to note that SR assumes the entities are already linked and is specifically designed for the Freebase knowledge graph, which ceased to be updated in 2015\footnote{\url{https://en.wikipedia.org/wiki/Freebase\_(database)}}.

\section{SRTK: Subgraph Retrieval Toolkit}

The SRTK project is publicly available on PyPI\footnote{\url{https://pypi.org/project/srtk}}, and it serves as both a command line toolkit and a Python library. We focus on demonstrating its usage as a command line toolkit. For information on using it as a Python library, please consult the Python API section of the documentation\footnote{\url{https://srtk.readthedocs.io}}. In this section, we first define the problem of subgraph retrieval, then we introduce each API and the underlying implementation. The SRTK CLI is divided into five subcommands:  \texttt{preprocess}, \texttt{train}, \texttt{link}, \texttt{retrieve}, and \texttt{visualize}. We categorize them based on two workflows: the first being retrieving subgraphs with trained models, and the second being training models for subgraph retrieval.

\subsection{Problem Definition}

The objective of \textbf{Semantic-relevant Subgraph Retrieval for KBQA} is to retrieve a subgraph $\mathcal{G} = (E_\mathcal{G}, R_\mathcal{G})$ from a given knowledge graph $G = (E, R)$ based on a natural language question $q$, such that $\mathcal{G} \subseteq G$ contains the answer entities or relations $A \subseteq G $ that the question expects. Here, $E$ and $R$ represent all entities and relations in the knowledge graph. The retrieved subgraph $\mathcal{G}$ is passed to a reasoner to predict answers. As reasoning is decoupled and not within the scope of this paper, we assume that the reasoner $p_\phi(A|q,\mathcal G)$ will select $m$ items from $\mathcal G$ as answers if it is known that there are $m$ answers.  In the best case, $\mathcal{G}$ will contain all the desired answers ($\mathcal{G} = A$), while in the worst case, none of the answers will be present in $\mathcal{G}$. If the answers included in $\mathcal{G}$ are fixed, a smaller subgraph $\mathcal{G}$ makes it easier for the reasoner $p_\phi(A|q,\mathcal G)$ to select the correct answers. On the other hand, when the size of $\mathcal{G}$ is fixed, including more answers in the subgraph increases the likelihood of the reasoner selecting the correct answers. Therefore, the goal of subgraph retrieval is to obtain a subset of the knowledge graph that is as close as possible to answers. The quality of the retrieval result is measured by both the size of the subgraph and the recall of the answers. One retrieved subgraph is considered strictly better than another if it has a smaller subgraph size while achieving a higher recall of the answers.

\subsection{Retrieval Workflow}

\begin{figure}
\centering
\includegraphics[width=\textwidth]{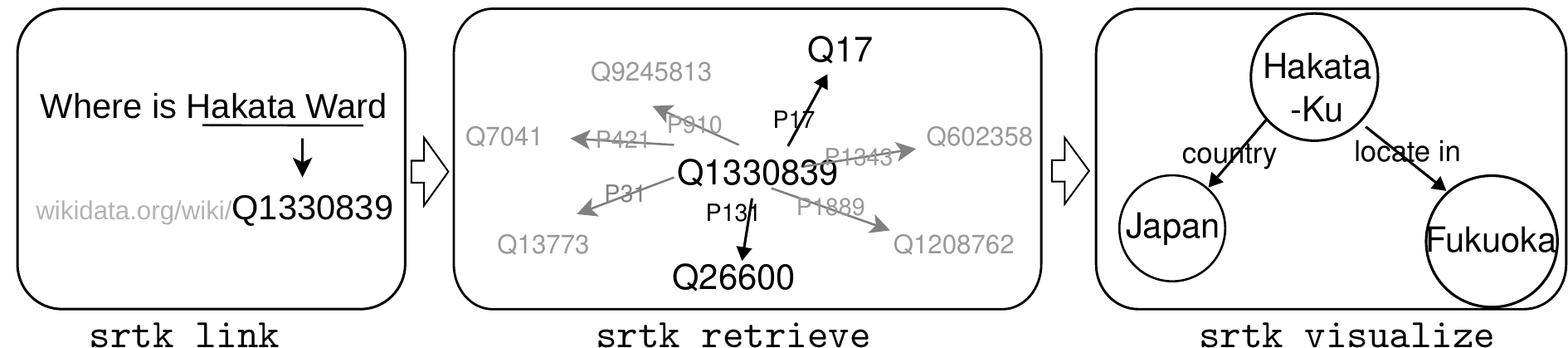}
\caption{The retrieval workflow of SRTK consists of three main steps: entity linking, subgraph retrieval, and visualization. Entity linking identifies knowledge graph entities mentioned in questions; subgraph retrieval retrieves semantic relevant subgraphs with trained models by iteratively including neighbors of the linked entities within certain proximity on a knowledge graph; visualization visualizes the retrieved subgraphs in the form of interactive webpages.} \label{retrieve-procedure}
\end{figure}

\subsubsection{Entity Linking}

Identifying named entities and aligning them with knowledge graph entities is typically the initial step in subgraph retrieval. If a dataset is already linked to a knowledge graph, this step may not be necessary.

SRTK \texttt{link} subcommand builds upon existing entity linking services, enabling researchers to perform entity linking on various knowledge graphs through the same interface. Consider the example \textit{Where is Hakata?}, once storing it in a jsonl file called \texttt{question.jsonl} like \texttt{\{"question": "Where is Hakata Ward?"\}}, users can link it to Wikidata with the following command:

\begin{lstlisting}[language=bash]
srtk link --input question.jsonl \
    --output linked.jsonl \
    --knowledge-graph wikidata \
    --el-endpoint https://rel-entity-linker.d4science.org
\end{lstlisting}

The link shown above is the public endpoint of REL \cite{vanHulst2020REL} service \footnote{In practice you also have to pass an authorization in the command line.}. Each line of the output \texttt{linked.jsonl} has the following format: 

\begin{lstlisting}[language=bash]
{"question_entities": ["Q1330839"], "spans": [[9,20]],
"entity_names": ["Hakata-ku,_Fukuoka"]}
\end{lstlisting}

Where the \textit{question entities} are the linked entities from Wikidata, the \textit{spans} store the corresponding character index of the linked entity in the original text, and the \textit{entity names} show the name of the linked entities in the knowledge graph. Under the hood, it first invokes REL \cite{vanHulst2020REL} to link entity mentions to Wikipedia articles, and then utilize \textit{wikimapper}\footnote{\url{https://github.com/jcklie/wikimapper}} to map the corresponding Wikipedia IDs to their corresponding entities in Wikidata. The extra step is needed because most existing services only link entities to Wikipedia \cite{vanHulst2020REL,ferragina2011tagme,piccinno2014wat}. 

For DBpedia, we integrate DBpedia Spotlight \cite{mendes2011dbpediaspotlight} to directly identify entity mentions and link them to DBpedia resources. By implementing annotation interfaces defined in SRTK, the entity linking can be extended to other knowledge graphs. 

\subsubsection{Retrieval}

The retrieval interface is at the core of the library. It retrieves a semantic-relevant subgraph with trained models and a given question.

The retrieval process can be divided into path search and fact retrieval. Path search identifies the most probable expansion paths from linked entities. An \textit{expansion path} comprises a sequence of relations from the knowledge graph, based on the idea that a question typically implies a reasoning chain \cite{zhang2022ruckbqa}. For the example \textit{Where is Hakata Ward?}, the corresponding expansion path would be \textit{<locate in>} from the triple \textit{(Hakata, locate in, ? )} implied by the question, where \textit{<locate in>} forms a path of length one. Path search starts with loading a trained path scoring model that measures the likelihood to expand a relation upon a current path, then it iteratively compares, selects, and includes neighboring relations into expansion paths. Path search initially regards linked entities and tracked entities. Subsequently, at each step of expansion, it first queries knowledge graphs to retrieve a set of relations connected to the tracked entities, then expand the paths by one hop by selecting the most probable relations with the trained scoring model, taking the question and the previously expanded paths into consideration. A path may also stop expansion when a special end relation is selected as the next relation. Afterward, path search updates the tracked entities with entities that the chosen relations connect to. During expansion, beam search is applied to keep track of top-$k$ most probable expansion paths; otherwise, the search space will grow explosively as hop increases. Finally, until a certain hop is reached, or until all paths have ended with the end relation, path search stops expansion. After getting a set of probable expansion paths, fact retrieval creates the subgraph by retrieving the entities and relations present along the paths.


The \texttt{retrieve} subcommand is used to retrieve subgraphs with linked questions and a trained scoring model. Users can additionally specify the max hop to search with \texttt{-{}-max-depth}, and beam width with \texttt{-{}-beam-width}. A larger beam width and a greater hop may increase the probability to arrive at target entities, but will also increase the running time. Using the \texttt{linked.jsonl} from the last sub-section with questions and linked entities, one can retrieve subgraphs with the following command:

\begin{lstlisting}[language=bash]
srtk retrieve --input linked.jsonl \
    --output subgraph.jsonl \
    --beam-width 2 --max-depth 1 \
    --scorer-model-path drt/srtk-scorer \
    --sparql-endpoint https://query.wikidata.org/sparql
\end{lstlisting}

The \texttt{-{}-scorer-model-path} specifies the location of a trained path scoring model. It can be either the folder of a saved local model or any language encoder model identifier from the Huggingface model hub\footnote{\url{https://huggingface.co/models}}. Besides, if an encoder-decoder model like T5 \cite{raffel2020t5} is fed in as the path scoring model, only the encoder part of the model will be used. The output has the following format, where the subgraph is saved as a list of \textit{(subject, predicate, object)} triples in the form of their identifiers on respective knowledge graphs.

\begin{lstlisting}[language=bash]
# Output subgraph.jsonl
{"triples": [["Q1330839","P31","Q26600"],
 ["Q1330839","P17","Q17"]]}
\end{lstlisting}

\subsubsection{Visualization}

SRTK's visualization module generates interactive web pages that display the retrieved subgraphs. The labels of the unique identifiers are fetched from the knowledge graphs through their SPARQL endpoints, and the linked entities are highlighted for better visualization. 

\begin{figure}
\centering
\includegraphics[width=0.5\textwidth]{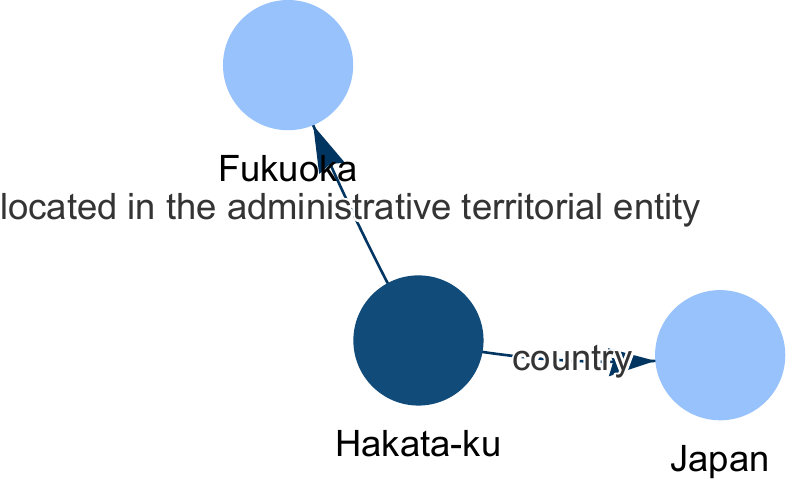}
\caption{The visualized subgraph with \texttt{srtk visualize}. The subgraph corresponds to the question \textit{Where is Hakata Ward?}. It is visualized from a subgraph comprised of two triples: (\textit{Hakata-ku, located in, Fukuoka}) and (\textit{Hakata-ku, country, Japan}).} \label{hakata-subgraph}
\end{figure}

By feeding the output subgraph \texttt{subgraph.jsonl} from the last sub-section, one can visualize it with the following command. The resulting graph is depicted in Fig.~\ref{hakata-subgraph}.

\begin{lstlisting}[language=bash]
srtk visualize --input subgraph.jsonl \
    --knowledge-graph wikidata \
    --sparql-endpoint https://query.wikidata.org/sparql
\end{lstlisting}

\subsection{Training Workflow}

SRTK trains scorer models to align questions with their corresponding reasoning paths in the embedding space, enabling semantic-relevant subgraph retrieval.

The training process can be achieved through either full supervision or weak supervision. In full supervision, the correct subgraph or the gold expansion paths are known. While in weak supervision, only the source and target entities are known, which is more common in KBQA. SRTK supports both supervised and weakly supervised learning. For the latter, SRTK searches for the shortest paths from source to target entities in the knowledge graphs during the preprocessing stage, which are then used as weak supervision signals.

\begin{figure}
\centering
\includegraphics[width=\textwidth]{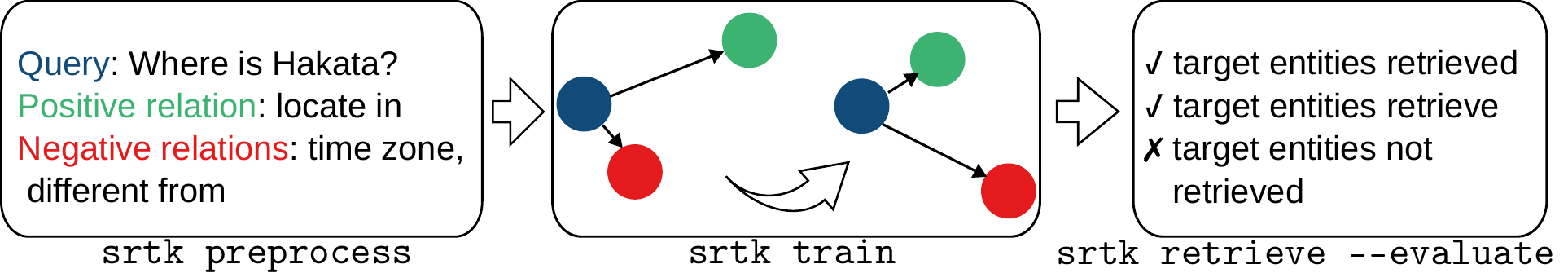}
\caption{The training workflow of SRTK involves three steps. Firstly, it preprocesses the raw data by identifying positive expansion paths and sampling negative ones. Next, it trains a scorer model that can selectively expand the subgraph according to a question by learning to pull together the question and the correct expansion path in the embedding space. Finally, if the answer entities are known, the scorer can be evaluated by calculating the answer coverage rate. The answer coverage rate in SRTK is calculated as the percentage of test samples where at least one answer entity is successfully retrieved.} \label{train-procedure}
\end{figure}

\subsubsection{Preprocessing}


The \texttt{preprocess} subcommand simplifies the preprocessing of training data. In the fully supervised scenario, where the expansion paths from source entities to target entities are known, the preprocessing create training samples from each expansion step within an expansion path. For weak supervision scenarios, where only the source and target entities are known, it additionally retrieves likely paths between source and target entities within two hops. Then it creates training samples as that in full supervision. Here, we discuss the case of weak supervision for completeness.

To extract the most likely paths, we first query the SPARQL endpoints to search for the shortest paths from each source entity to the target entities. Since there may be multiple paths between the source and target entities, we then filter the paths by examining how close the entity sets derived from the paths to the answers. This involves retrieving a set of terminal entities of each path and scoring the paths based on the Jaccard index between the retrieved entities and the answer entities. Paths with scores below a certain threshold are discarded.

Once the expansion paths are known, we convert the retrieved paths into training samples by sampling negative relations for each relation in an expansion path. The intuition of this conversion is to mimic the decision-making process during retrieval, where the goal is to select the most suitable relation from the set of connected relations so that the probability of reaching the target entities will be maximized. This is a form of weak supervision, as the retrieved expansion path may not be the real reasoning chain. A $K$-hop path is decomposed into $K+1$ positive samples. The first $K$ samples are $([q; r_1, r_2, \ldots, \allowbreak r_{k-1}], r_k)$, with $q$ being a given query, , $r_k$ being the $k$-th relation, and $k$ in range $1,2,\ldots,K$. The last sample is $([q; r_1,\ldots, r_K]; \textnormal{END})$, where $\textnormal{END}$ is a special relation signifying stopping expansion.  For a positive sample $([q; r_1, r_2, \ldots, r_{k-1}], r_k)$, $N$ negative samples are retrieved from the knowledge graph by sampling from the relations connected to relation $r_{k-1}$ via one-hop intermediate entities, excluding $r_k$. Each positive sample is then converted to a training sample along with its negative samples and the query.

For instance, if we know from the previous example that the source entity is \texttt{Q1330839} (Hakata-ku), while the target entity is \texttt{Q26600} (Fukuoka City), we can first arrange the information into a JSONL file with one line as follows:

\begin{lstlisting}[language=bash]
# Input kbqa_dataset.jsonl
{"question": "Where is Hakata Ward?",
 "question_entities": ["Q1330839"],
 "answer_entities": ["Q26600"]}
\end{lstlisting}

By executing the \texttt{srtk preprocess} command as follows, we first identify the probable expansion paths between the question and answer entities within two hops, then create training samples from the paths:

\begin{lstlisting}[language=bash]
srtk preprocess --input kbqa_dataset.jsonl \
    --search-path --output train.jsonl \
    --knowledge-graph wikidata \
    --sparql-endpoint https://query.wikidata.org/sparql \
    --metric jaccard --num-negative 2
\end{lstlisting}

In the command above, \texttt{-{}-search\allowbreak-path} instructs SRTK to look for connections between question and answer entities. This argument can be omitted when the gold expansion paths are known. \texttt{-{}-num-negative} specifies how many negative relations to sample for each relation. The first sample in \texttt{train.jsonl} is as follows:

\begin{lstlisting}[language=bash]
{"query": "Where is Hakata Ward? [SEP]",
 "positive": "located in the administrative entity",
 "negatives": ["located in time zone","different from"]}
\end{lstlisting}

In the output, the \textit{query} is the question and previous expansion relations concatenated by a $\textnormal{[SEP]}$ token. The \textit{positive} field saves the relation between Hakata-ku and Fukuoka as natural language, while the \textit{negative} field saves a list of possible negative relations that connects to Hakata-ku but are not in the correct expansion path. Since the question and answers are only one hop away in this case, there will be only two generated training samples, where the first is as shown above, having the one-hop relation as a positive relation, while the second has $\textnormal{END}$ as the positive relation.

\subsubsection{Training and Evaluation}


The training process aims to train an encoder that encodes questions close to their correct expansion paths in the embedding space. To be more specific, an encoder is trained to encode each relation on a correct expansion path close to the question concatenated with all relations before that relation. For example, if a question $q$ has the reasoning path $r1 \rightarrow r2$, then an encoder is trained to encode $r1$, $r2$, and the special relation $\textnormal{END}$ similar with $q$, $q + r1$, and $q + r1 + r2$, respectively. Here, $+$ denotes concatenation. We call a trained encoder a \textit{scorer}, as it scorers candidate relations at inference time. Therefore, for the question \textit{Where is Hakata?} and the expansion path \textit{<locate in>}, \textit{Where is Hakata? [SEP]} and \textit{locate in} should be similar if measured by a scorer; the same is for \textit{Where is Hakata? [SEP] locate in} and $\textnormal{END}$. The similarity is computed as the cosine distance of the sentence-level embeddings. For BERT-like models \cite{devlin2019bert}, which use a $[\textnormal{CLS}]$ token for classification, we use the embedding of $\textnormal{[CLS]}$ as the sentence-level embedding. For other encoder models, we average over all token embeddings in the last hidden layer of a sentence to obtain its sentence-level embedding.


We finetune a pretrained language encoder from huggingface as a scorer. Two loss functions for model optimization are supported: cross-entropy loss and contrastive loss. In the cross-entropy loss, the positive relation is regarded as the correct class, while the other $N$ negative relations are regarded as false ones. For contrastive learning, we use NTXentLoss \cite{chen2020ntxentloss}, where the question and the positive relation form a positive pair, and the query and each of the negative relation form negative pairs.

Consider the output \texttt{train.jsonl} from the previous sub-section, if there are many training samples generated as before, we can train a scorer with \texttt{srtk train} as follows:

\begin{lstlisting}[language=bash]
srtk train --train-dataset data/train.jsonl \
    --output-dir artifacts/scorer \
    --model-name-or-path roberta-base \
    --accelerator gpu
\end{lstlisting}


If the answer entities are known in a dataset, we can proceed to evaluate a scorer using SRTK. It works by retrieving subgraphs for a test dataset, then calculating how many answer entities are retrieved with answer coverage rate. We define answer coverage rate as the number of samples where any of the correct answer entities is retrieved, divided by the total number of test samples. We can perform an evaluation using the \texttt{srtk retrieve} subcommand with the \texttt{-{}-evaluate} option. The expected input fields include question, question entities, and answer entities. If the test dataset is stored at \texttt{test.jsonl}, one can evaluate a trained scorer with the following command:

\begin{lstlisting}[language=bash]
srtk retrieve --input test.jsonl \
    --scorer-model-path artifacts/scorer \
    --evaluate
>>> Answer coverage rate: 0.9749 (4188 / 4296)
>>> Average subgraph size: 7.5345 triples
\end{lstlisting}

In the provided example, the command line output reveals that out of the 4296 samples, a total of 4188 samples successfully retrieve at least one of the target entities. Additionally, the size of the subgraphs is measured by calculating the average number of triples within the subgraphs across all the samples. The numbers provided in this example are used purely for explanatory purposes.

\section{Discussions}

\subsection{Impact}

In this paper, we have primarily focused on discussing the use case of SRTK in the domain of KBQA. Within KBQA, our toolkit helps researchers more easily retrieve subgraphs, which reduces search space and may improve the downstream tasks by providing less noisy information. Moreover, SRTK promotes the migration from outdated knowledge graphs to up-to-date ones by providing unified interfaces. For example, research works that use to depend on Freebase can seamlessly migrate to Wikidata\footnote{We additionally provide migration scripts from Freebase to Wikidata, including entities and relations: \url{https://github.com/happen2me/freebase-wikidata-convert}} by updating the knowledge graph endpoint URL, while the interfaces to access underlying knowledge graphs remain the same.  More importantly, SRTK may promote the development of transferable and shareable\footnote{The trained models can be easily shared on HuggingFace hub, e.g. we share our trained checkpoint as \texttt{drt/srtk-scorer}} retrieval algorithms, such that newly developed datasets or algorithms work across knowledge graphs and can be easier compared and benchmarked.

SRTK has numerous potential use cases that extend beyond KBQA. In knowledge graph augmented language model pretraining, subgraphs retrieved with SRTK can be used to enhance language representations with either entity knowledge \cite{zhang2019ernie,yamada2020luke,xiong2019encyclopedia,peters2019knowbert,fevry2020eae} or triple knowledge \cite{sun2020colake,wang2021kepler,qin2021erica}. One possible approach is using SRTK to identify entities, search relations within the linked entities or with semantic-relevant neighbors, and use the retrieved triples for language pretraining. In knowledge-enhanced language generation, SRTK may be used to retrieve relevant and accurate facts with the given prompts \cite{koncel2019textgenfromkg,ji2020languagegen,liu2020kbert,zhou2019improving}. In conversation reasoning and generation, SRTK can be used to identify mentioned entities and provides semantic-relevant subgraphs as extra information for responses\cite{zhang2020grounded,moon2019opendialkg,tuan2019dykgchat}. In fact verification, SRTK may be used to retrieve subgraph as reliable facts to verify the statements \cite{cao2018faithful}. One way to accomplish this is to examine whether certain links exist in the retrieved subgraph. Furthermore, the subgraphs retrieved by SRTK can be leveraged for improvement on a variety of downstream tasks \cite{yu2022surveyketg,zhen2022survey}, like translation \cite{moussallem2019thoth}, summarization \cite{huang2020summarize} etc.

\subsection{Positioning to the State of the Art}

SRTK stands on the shoulder of giants. It builds upon existing entity linking services and extends an existing state-of-the-art path expansion algorithm to offer an off-the-shelf toolkit for subgraph retrieval across various knowledge graphs. Regarding entity linking on Wikidata, REL \cite{vanHulst2020REL} is integrated, which currently achieves the highest micro-F1-strong score on the AIDA CoNLL-YAGO dataset \cite{hoffart2011aidadataset}. Although REL and most other entity linking tools primarily link entities to Wikipedia, the presence of a maintained mapping between Wikipedia articles and Wikidata enables their utilization. For entity linking on DBpedia, DBpedia Spotlight \cite{mendes2011dbpediaspotlight} is supported. It relies on pattern match and thus does not perform as well as neural models \cite{sevgili2022elsurvey}. However, due to the absence of alternative off-the-shelf entity linking services for DBpedia, we include DBpedia Spotlight as part of the toolkit.  For path expansion, we integrate SR proposed by Zhang et al \cite{zhang2022ruckbqa}. They reported state-of-the-art results on CWQ \cite{talmor2018cwq} and WebQSP \cite{yih2016webqsp}, two datasets for KBQA on Freebase, with the subgraph retrieved by SR. The retrieved subgraphs are significantly smaller in size but still have high coverage of answer entities. We further extend SR to support other knowledge graphs. Although we did not compare the results of KBQA using subgraphs retrieved by SRTK on datasets of other knowledge graphs, this is due to the complexities involved in setting up such comparisons. Besides, evaluating the performance of a specific implementation of the path expansion algorithm in various settings exceeds the scope of this paper.

\subsection{Limitations}

SRTK relies on knowledge graph endpoints for up-to-date information, but network latency and iterative queries slow down retrieval. To address this, we propose two solutions. Firstly, setting up local endpoints can alleviate the problem, as demonstrated in our experiments (tutorial available in our documentation). Secondly, caching known entities' $k$-hop facts in advance can build a smaller local graph, an approach currently under development.

Another limitation of SRTK is its dependency on preceding steps, where issues can accumulate and magnify. For instance, failure to link entities to knowledge graphs renders subgraph retrieval impossible. One possible solution is to combine the results from multiple entity linking services or to fallback to n-gram based methods when neural methods fail so that the mentioned are more likely to be recognized. We plan to integrate more existing entity linking services into SRTK.

On the path expansion side, there are two notable limitations. One limitation is the restricted expansion direction. SRTK currently only expands outward along directed relations. This means for a triple like \textit{(Hakata, locate in, Fukuoka)}, \textit{Fukuoka} can be discovered by following the directed relation \textit{locate in} if \textit{Hakata} is known, but the inversed discovery of \textit{Hakata} is not possible if only \textit{Fukuoka} is known. This limitation arises in knowledge bases lacking paired inverse relations like Wikidata. Another limitation lies in the proximity assumption. The current link-then-expand approach assumes that the target entities lie in a certain proximity to the linked entities. This holds for most KBQA scenarios but does not generally hold for all subgraph retrieval situations. The first limitation can be alleviated by allowing inverse expansion, e.g. constructing SPARQL queries with unknown variables in the subject position. But this significantly increases complexity, including challenges of circle avoidance and expansion direction determination. The second limitation may be solved by replacing entity linking with problem-specific entity discovery methods to locate entities near the target.

Besides, SRTK currently retrieves subgraphs as sets of triples, excluding additional affiliated information available in knowledge graphs like Wikidata. For example, the triple \textit{(Merkel, Position Held, Chancellor of Germany)} includes start and end dates. Current methods retrieve this information as extra triples (position held, start time, 22.11.2005). Future improved algorithms seek to retrieve all related information simultaneously.

\subsection{Future Development and Maintenance}

The SRTK project has several directions for future development. In addition to the current capabilities, we plan to add support for additional knowledge graphs such as YAGO \cite{pellissier2020yago}. This would involve integrating an entity linking service and implementing graph access interfaces for each supported knowledge graph. Besides, since the abstract access interfaces are implementation-agnostic, SRTK has the potential to support knowledge graphs queried with interfaces other than SPARQL. Our ultimate goal is to encourage comparable methods across different knowledge graphs and to help researchers migrate from outdated knowledge graphs.

Moving forward, we aim to expand the range of algorithms supported by SRTK. The current retrieval algorithm is path-centric: it retrieves subgraphs by selecting the most probable outgoing expansion paths. However, we see room for improvement in several areas. For example, we could take entities along the paths into account when expanding subgraphs, and consider both incoming and outgoing links when expanding from a known entity. This would undoubtedly make expansion more complex, but it holds promise for future research.

In the medium term, our plans for SRTK include incorporating support for YAGO, integrating more online entity linking services, and exploring the compatibility of SRTK with existing information retrieval-based KBQA methods across various datasets. Looking ahead to the long term, we aspire to incorporate support for additional retrieval algorithms and integrate SRTK with popular KBQA models, enabling direct comparisons and benchmarking of existing KBQA and retrieval methods. Additionally, we are also committed to continuously maintaining SRTK, addressing any bugs or security vulnerabilities. To achieve these visions, we also greatly value and encourage contributions from the open-source community.

\section{Conclusion}

SRTK is an extensible and user-friendly toolkit that focuses on retrieving semantic-relevant subgraphs. It is built on a state-of-the-art algorithm and comes with a command line interface and Python API. SRTK streamlines the full life cycle of subgraph retrieval development and application, providing customizable pipeline steps. It supports multiple knowledge graphs, including Freebase, DBpedia, and Wikidata. Future developments include support for entity linking services, more knowledge graphs, and other retrieval algorithms.

\bigskip

\noindent \textit{Resource Availability Statement}: Source code for SRTK is available from Github\footnote{\url{https://github.com/happen2me/subgraph-retrieval-toolkit}}. Its canonical citation is \cite{shen2023srtk}.

%
%
\bibliographystyle{splncs04}
\bibliography{main}

\end{document}